\documentclass[aps,prl,twocolumn,superscriptaddress,showpacs,floatfix]{revtex4}

\newcommand{\St}{\mathcal{S}}
\usepackage{graphicx}
\usepackage{bm}

\begin{document}

\title{Heavy particle concentration in turbulence at dissipative and
  inertial scales}

\author{J.\ Bec} \affiliation{CNRS UMR6202, Observatoire de la C\^ote
  d'Azur, BP4229, 06304 Nice Cedex 4, France.}

\author{L.\ Biferale} \affiliation{Dip.\ Fisica and INFN, Universit\`a
  ``Tor Vergata'', Via Ricerca Scientifica~1, 00133 Roma, Italy.}

\author{M.\ Cencini} \affiliation{INFM-CNR, SMC Dept.\ of Physics,
  Universit\`a ``La Sapienza", P.zzle A.~Moro~2, 00185 Roma, Italy,
  and\\ CNR-ISC, Via dei Taurini 19, 00185 Roma, Italy.}

\author{A.\ Lanotte} \affiliation{CNR-ISAC and INFN, Sezione di Lecce, Str.\
  Prov.\ Lecce-Monteroni, 73100 Lecce, Italy.}

\author{S.\ Musacchio} \affiliation{Weizmann Institute of Science,
Department of Complex Systems, 76100 Rehovot, Israel.}

\author{F.\ Toschi} \affiliation{CNR-IAC, Viale del Policlinico 137,
  00161 Roma, Italy, and\\ INFN, Sezione di Ferrara, Via G.\
  Saragat 1, 44100 Ferrara, Italy.}

\begin{abstract}
  Spatial distributions of heavy particles suspended in an
  incompressible isotropic and homogeneous turbulent flow are
  investigated by means of high resolution direct numerical
  simulations. In the dissipative range, it is shown that particles
  form fractal clusters with properties independent of the Reynolds
  number. Clustering is there optimal when the particle response time
  is of the order of the Kolmogorov time scale $\tau_\eta$.  In the
  inertial range, the particle distribution is no longer
  scale-invariant. It is however shown that deviations from uniformity
  depend on a rescaled contraction rate, which is different from the
  local Stokes number given by dimensional analysis.  Particle
  distribution is characterized by voids spanning all scales of the
  turbulent flow; their signature in the coarse-grained mass
  probability distribution is an algebraic behavior at small
  densities.
\end{abstract}
\date{\today}

\maketitle

Understanding the spatial distribution of finite-size massive
impurities, such as droplets, dust or bubbles suspended in
incompressible flows is a crucial issue in engineering~\cite{cst98},
planetology~\cite{pl01} and cloud physics~\cite{ks01}. Such particles
possess inertia, and generally distribute in a strongly inhomogeneous
manner.  The common understanding of this long known but remarkable
phenomenon of \emph{preferential concentrations} relies on the idea
that, in a turbulent flow, vortexes act as centrifuges ejecting
particles heavier than the fluid and entrapping lighter
ones~\cite{ef94}.  This picture was successfully used (see, e.g.\
\cite{ef94,ck04}) to describe the small-scale particle distribution
and to show that it depends only on the Stokes number
$\St_\eta=\tau/\tau_\eta$, which is obtained by non-dimensionalizing
the particle response time $\tau$ with the characteristic time
$\tau_\eta$ of the small turbulent eddies.

In this Letter, we confirm that this mechanism is relevant to describe
the particle distribution at length-scales which are smaller than the
dissipative scale of turbulence, $\eta$. In particular, maximal
clustering is found for Stokes numbers of the order of unity. However,
we show that stationary particle concentration experiences also very
strong fluctuations in the inertial range of turbulence. In analogy
with small-scale clustering, it is expected that for $r\gg \eta$ the
relevant parameter is the local Stokes number $\St_r=\tau/\tau_r$,
where $\tau_r$ is the characteristic time of eddies of size
$r$~\cite{ffs03}. Surprisingly, we present evidences that such
a dimensional argument does not apply to describe how particles organize
in the inertial range of turbulence. We show that the way they
distribute depends on a scale-dependent rate at which volumes are
contracted.

In very dilute suspensions, the trajectory $\bm X(t)$ of small
spherical particles much heavier than the fluid evolve according to
the Newton equation~\cite{m87}
\begin{equation}
  \tau \ddot{\bm X} ={\bm u(\bm X,t) - \dot{\bm X}}\,.
  \label{eq:newton}
\end{equation}
The response time $\tau$ is proportional to the square of the
particles size and to their density contrast with the fluid. Here, we
neglect buoyancy and particle Brownian diffusion. As stressed
in~\cite{bff01,eklrs02}, diffusion may affect concentration of
particles. However we assume that the typical lengthscale below which
particle diffusion becomes dominant is much smaller than the
Kolmogorov scale $\eta$ of the fluid flow, and than any observation
scale considered here. In most situations, particles are so massive
that such an approximation is fully justified.

The fluid velocity $\bm u$ satisfies the incompressible Navier-Stokes
equation
\begin{equation}
  \partial_t\bm u+ \bm u\cdot\bm \nabla\bm u=-\bm \nabla
  p+\nu\nabla^2\bm u+\bm F \;\; {\mbox {with}} \;\; \bm \nabla \cdot
  \bm u=0\,,
\label{eq:ns}
\end{equation}
$p$ being the pressure and $\nu$ the viscosity; $\bm F$ is an external
homogeneous and isotropic force that injects energy at large scales
$L$ with a rate $\varepsilon=\langle\bm F\cdot\bm u \rangle$.  We
performed direct numerical simulations of (\ref{eq:ns}) by means of a
pseudo-spectral code on a triply periodic box of size
$\mathcal{L}=2\pi$ with $128^3$, $256^3$ and $512^3$ collocation
points corresponding to Reynolds numbers (at the Taylor micro-scale)
$R_\lambda \approx 65,\,105$ and $185$, respectively. Once the flow
$\bm u$ is statistically stationary, particles with 15 different
values of the Stokes number in the range $\St_\eta \in [0.16,3.5]$
($N\!=\!7.5$ millions of particles per $\St_\eta$), are seeded
homogeneously in space with velocity equal to that of the fluid, and
are evolved according to (\ref{eq:newton}) for about two large-scale
eddy turnover times. After this time, particle mass distribution
reaches a statistical steady state too. Measurements are then
performed. Details on the numerics and on the transient are reported in
\cite{noi1}.

\begin{figure}[t!]
  \centerline{\includegraphics[width=0.4\textwidth]{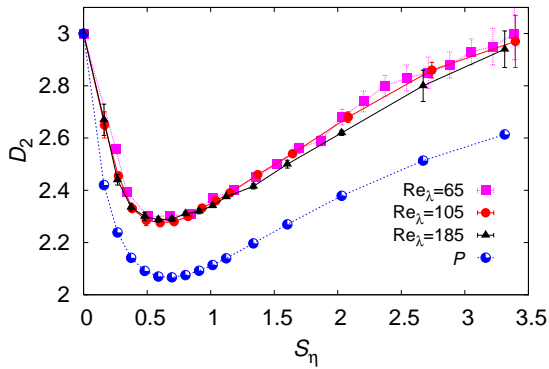}}
  \caption{(Color online) The correlation dimension $\mathcal{D}_2$ vs
    $\St_\eta$ for the three different $R_\lambda$. Also shown the
    probability $\mathcal{P}$ to find particles in non-hyperbolic
    (rotating) regions of the flow, for $Re_\lambda=185$ (multiplied
    by an arbitrary factor for plotting purposes).  $\mathcal{D}_2$
    has been estimated taking into account also subleading terms, as
    described in~\protect\cite{bch06}.}
  \label{correldim}
\end{figure}
Below the Kolmogorov length-scale $\eta$ where the velocity field is
differentiable, the motion of inertial particles is governed by the
fluid strain and the dissipative dynamics leads their trajectories to
converge to a dynamically evolving attractor. For any given response
time of the particles, their mass distribution is singular and
generically scale-invariant with fractal properties at small
scales~\cite{ekr96,bff01}. In order to characterize particle clusters
at these scales, we measured the correlation dimension
$\mathcal{D}_2$, which is estimated through the small-scale algebraic
behavior of the probability to find two particles at a distance less
than a given $r$: $P_2(r)\sim r^{\mathcal{D}_2}$.  The dependence of
$\mathcal{D}_2$ on $\St_\eta$ and $Re_\lambda$ is shown in
Fig.~\ref{correldim}. Notice that $\mathcal{D}_2$ depends very weakly
on $Re_\lambda$ in the range of Reynolds numbers here explored. A
similar observation was done in ~\cite{ck04}, where particle
clustering was equivalently characterized in terms of the radial
distribution function.  This indicates that $\tau_\eta$, which varies
by more than a factor 2 between the smallest and the largest Reynolds
numbers considered here, is the relevant time scale to characterize
clustering below the Kolmogorov scale $\eta$.  For all values of
$Re_\lambda$, a maximum of clustering (minimum of $\mathcal{D}_2$) is
observed for $\St_\eta\approx 0.6$.  For values of $\St_\eta$ larger
than those investigated here, $\mathcal{D}_2$ is expected to saturate
to the space dimension~\cite{bch06}. For small values of $\St_\eta$,
particle positions strongly correlate with the local structure of the
fluid velocity field. A quantitative measure is also given in
Fig.~\ref{correldim} by the probability $\mathcal{P}$ to find
particles in non-hyperbolic regions of the flow, i.e.\ at those points
where the strain matrix has two complex conjugate eigenvalues.
\begin{figure}[t]
  \centerline{\includegraphics[width=0.3\textwidth]{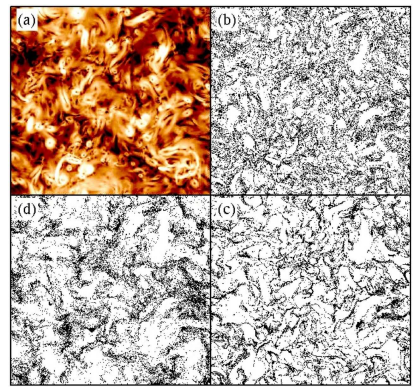}}
 \vspace{-3pt}
  \caption{(Color online) (a) The modulus of the pressure gradient,
    giving the main contribution to fluid acceleration, on a slice
    $512\!\times\! 512\!\times\! 4$. B/W code low and high intensity,
    respectively.  Particle positions in the same slice are shown for
    (b) $\St_\eta\!=\!0.16$, (c) $0.80$ and (d) $3.30$. Note the
    presence of voids with sizes much larger than the dissipative
    scale.}
  \label{snapshot}
\end{figure}
Note that $\mathcal{P}$ attains its minimum for values of $\St_\eta$
close to the minimum of $\mathcal{D}_2$. This supports the traditional
view relating particle clustering to vortex ejection. As qualitatively
evidenced from Fig.~\ref{snapshot}, particle positions correlate with
low values of the fluid acceleration. This phenomenon was already
evidenced in \cite{noi1} where a statistical analysis of the fluid
acceleration at particle positions was performed, and also
in~\cite{cgv06} for 2D turbulent flows.

From Fig.~\ref{snapshot}, it is clear that fluctuations in the
particle spatial distribution extend to scales far inside the inertial
range; this confirms the experimental measurements
of~\cite{achl02}. Moreover, for Stokes numbers of the order of unity
(Figs.~\ref{snapshot}c and \ref{snapshot}d), we note that the sizes of
voids span all spatial scales of the fluid turbulent flow, similarly
to what observed in 2D inverse cascade turbulence~\cite{bdg04,cgv06}.
To gain a quantitative insight, we consider the Probability Density
Function (PDF) $P_{r,\tau}(\rho)$ of the particle density
coarse-grained on a scale $r$ inside the inertial range, that is the
probability distribution of the fraction of particles in a cube of
size $r$, normalized by the volume $r^3$ of the cube. For tracers,
which are uniformly distributed, and for an infinite number of
particles $N \to \infty$, this PDF is a delta function on $\rho = 1$.
For finite $N$, the probability to have a number $n$ of particles in a
box of size $r$ is given by the binomial distribution and leads to a
closed form for the PDF of the coarse-grained density $\rho =
n\mathcal{L}^3/(Nr^3)$.  In our settings, the typical number of
particles in cells inside the inertial range is several thousands.
Thus, finite number effects are not expected to affect the core of the
mass distribution $\rho\! = \!\mathcal{O}(1)$, but they may be
particularly severe when $\rho\ll 1$ because of the presence of
voids. To reduce this spurious effect, we computed the
quasi-Lagrangian (QL) mass density PDF $P^{(QL)}_{r,\tau}(\rho)$,
which corresponds to a Lagrangian average with respect to the natural
measure~\cite{bgh04}, and is obtained by weighting each cell with the
mass it contains.  For statistically homogeneous distribution, QL
statistics can be related to the Eulerian ones by noticing that
$\langle \rho_r^p\rangle_{QL} = \langle \rho_r^{p+1} \rangle$
(see, e.g.,~\cite{bgh04}).

\begin{figure}[t!]
  \centerline{\includegraphics[width=0.47\textwidth]{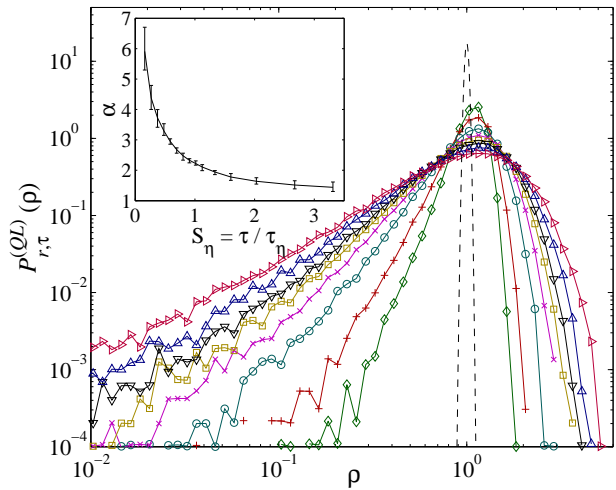}}
  \vspace{-12pt}
  \caption{(Color online) The quasi-Lagrangian PDF of the
    coarse-grained mass density $\rho_r$ in log-log scale for
    $\St_\eta=0.27$, $0.37$, $0.58$, $0.80$, $1.0$, $1.33$, $2.03$,
    $3.31$ (from bottom to top) at scale $r =
    \mathcal{L}/16=0.39$. The dashed line represents a uniform
    distribution.  Inset: exponent of the power law left tail $\alpha$
    vs $\St_\eta$. It has been estimated from a best fit of the
    cumulative probability which is far less noisy. Data refer to
    $Re_\lambda=185$.}
  \label{pdf_rho_QLag}
\end{figure}

Figure~\ref{pdf_rho_QLag} shows $P^{(QL)}_{r,\tau}(\rho)$, for various
response times $\tau$ and at a given scale ($r=\mathcal{L}/16$) within
the inertial range.  Deviations from a uniform distribution can be
clearly observed, they become stronger and stronger as $\tau$
increases.  This means that concentration fluctuations are important
not only at dissipative scales but also in the inertial range.  A
noticeable observation is that the low-density tail of the PDF (which
is related to voids) displays an algebraic behavior
$P^{(QL)}_{r,\tau}(\rho)\sim \rho^{\alpha(r,\tau)}$. This means that
the Eulerian PDF of the coarse-grained mass density has also a
power-law tail for $\rho\ll 1$, with exponent $\alpha-1$.  The
dependence of this exponent for fixed $r$ and varying the Stokes
number $\mathcal{S}_\eta$ is shown in the inset.  For low inertia
($\mathcal{S}_\eta\to0$), it tends to infinity in order to recover the
non-algebraic behavior of tracers. At the largest available Stokes
numbers, we observe $\alpha \to 1$, indicating a non-zero
probability for totally empty areas.  Note however that $\alpha$ is
expected to go to infinity when $\mathcal{S}_\eta\to\infty$ with $r$
fixed, since a uniform distribution is expected for infinite
inertia. Algebraic tails are relevant to various physical/chemical
processes involving heavy particles.  The particle distribution in low
density regions is an important effect to be accounted for, e.g.,
modelling the growth of liquid droplets by condensation of vapor onto
aerosol particles~\cite{bff01} and of aerosol scavenging~\cite{ks01}.

Fixing the response time $\tau$ and increasing the observation scale
$r$ reproduces the same qualitative picture as fixing $r$ and
decreasing $\tau$.  A uniform distribution is recovered in both limits
$r\to\infty$ or $\tau\to 0$.  These two limits are actually
equivalent.  At length-scales $r\gg\eta$ within the inertial range,
the fluid velocity field is not smooth: according to Kolmogorov 1941
theory (K41), velocity increments behave as $\delta_r u \sim
(\varepsilon r)^{1/3}$.  K41 theory suggests that fluctuations at
scale $r$ are associated to time scales of the order of the turnover
time $\tau_r = r/\delta_r u\sim \varepsilon^{-1/3}r^{2/3}$. This
implies that for any finite particle response time $\tau$ and at
sufficiently large scales $r$, the local inertia measured by
$\St_r=\tau/\tau_r$ becomes so small that particles should behave as
tracers and distribute uniformly in space~\cite{ffs03}.  Deviations
from uniformity for finite $\St_r$ are expected not to be
scale-invariant~\cite{bff01}. In particular, observations from random
$\delta$-correlated in time flows \cite{bch06b} suggest that particle
distribution should depend only on the local Stokes number
$\St_r$. However, as explained below, this argument does not seem to
apply to realistic flows.

The presence of inhomogeneities in the spatial distribution of
particles is due to a dynamical competition between stretching,
folding, and contraction due to their dissipative dynamics. At small
scales where the flow is spatially smooth, these mechanisms are
described by the Lyapunov exponents and their finite-time
fluctuations, associated to the growth and contraction rates of
infinitesimal volumes. At larger scales, competition between particle
spreading and concentration is also at equilibrium.  The spatial
distribution of particle at a given scale $r\gg\eta$ should only
depend on the time scale given by the inverse of the contraction rate
$\gamma_{r,\tau}$ of a size-$r$ blob $\mathcal{B}_r$ of particles with
response time $\tau$. However in the inertial range, the flow is not
differentiable and an approach based on Lyapunov exponents cannot be
used. In particular, the rate $\gamma_{r,\tau}$ should depend on
$r$. To estimate this contraction rate in the limit of small $\tau$,
we make use of Maxey's approximation~\footnote{For describing scales
  $r>\eta$ the approximation is expected to be valid whenever
  $\St_r\ll 1$, i.e. for essentially all the explored range of
  $\tau$'s.} of the particle dynamics~\cite{m87}
\begin{equation}
  \dot{\bm X} \approx \bm v(\bm X,t)\, \;\; {\mbox {with}} \;\; \bm v
  = \bm u -\tau \left(\partial_t \bm u + \bm u\cdot\nabla\bm
  u\right)\,,
  \label{eq:maxeyapprox}
\end{equation}
meaning that particle trajectories can be approximated by those of
tracers in the synthetic compressible velocity field $\bm v$. The
inertial correction is proportional to the fluid acceleration that, in
turbulent flows, is itself dominated by pressure gradient.  The
contraction rate of the blob $\mathcal{B}_r$ with volume $r^3$ is
$\gamma_{r,\tau}= (1/r^3)\! \int_{\mathcal{B}_r} \!\! d^3x\, \nabla
\!\cdot\!  {\bm v}(\bm x)$. As $\nabla \!\cdot\! {\bm v} \!\simeq\!
\tau \nabla^2 p$, we have $\gamma_{r,\tau} \sim (\tau / r^2)\,
\delta_r p$, where $\delta_r p$ denotes the typical pressure increment
at scale $r$. This means that the scale dependence of the contraction
rate is directly related to the scaling properties of the pressure
field. Dimensional arguments suggest that $\delta_r p \sim
(\varepsilon r)^{2/3}$, so that the contraction rate scales as
$\gamma_{r,\tau} \sim \tau \varepsilon^{2/3} / r^{4/3}$. However, as
stressed in~\cite{gf01-ti03}, K41 scaling for pressure is observable
only when the Reynolds number is tremendously large (typically for
$Re_\lambda \gtrsim 600$). In our simulations, where $Re_\lambda < 200
$, pressure scaling is actually dominated by random
sweeping~\cite{t75}, and we observe $\delta_r p \sim U\,(\varepsilon
r)^{1/3}$ (as in other simulations at comparable Reynolds,
see~\cite{gf01-ti03}). This implies that the contraction rate is
$\gamma_{r,\tau} \sim \tau \varepsilon^{1/3} U / r^{5/3}$.

\begin{figure}[t!]
  \centerline{\includegraphics[width=0.5\textwidth]{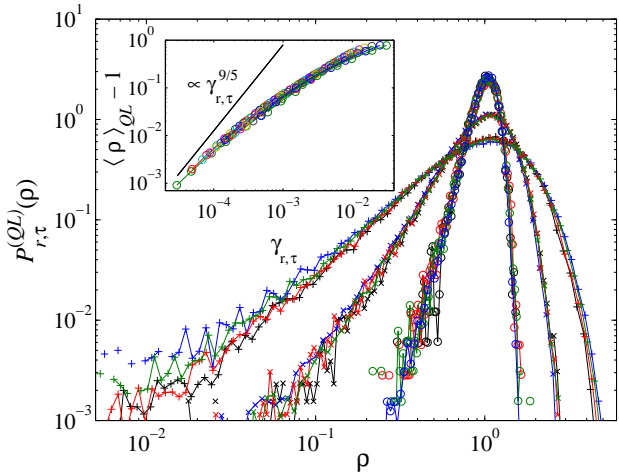}}
  \vspace{-12pt}
  \caption{(Color online) Pdf of the coarse-grained mass in the
    inertial range for three values of the non-dimensional contraction
    rate $\gamma_{r,\tau} \tau_\eta$.  From bottom to top:
    $\gamma_{r,\tau}=4.8\times 10^{-4}/\tau_\eta$ (different curves
    refer to $\St_\eta = 0.16$, $0.27$, $0.37$, $0.48$),
    $\gamma_{r,\tau}=2.1\times 10^{-3}/\tau_\eta$ (for $\St_\eta
    =0.58$, $0.69$, $0.80$, $0.91$, $1.0$), $\gamma_{r,\tau}=7.9\times
    10^{-3}/\tau_\eta$ (for $\St_\eta =1.60$, $2.03$, $2.67$,
    $3.31$). Inset: deviation from unity of the first-order QL moment
    for $r$ within the inertial range.  For comparison, the behavior
    $\propto\gamma_{r,\tau}^{9/5}$ obtained when assuming uniformly
    distributed pointlike clusters of particles is shown as a solid
    line. Data refer to $Re_\lambda=185$.}
  \label{collapseQL}
\end{figure}

Figure~\ref{collapseQL} shows $P_{r,\tau}^{(QL)}(\rho)$ for three
choices of $\gamma_{r,\tau}$ obtained from various sets of values of
$r$ and $\tau$. The collapse of the different curves strongly supports
that the distribution of the coarse-grained mass density depends only
upon the scale-dependent contraction rate $\gamma_{r,\tau}$. In
particular, as represented in the inset of Fig.~\ref{collapseQL}, the
deviations from unity of the first moment of the distribution collapse
for all $\St_\eta$ investigated and all scales inside the inertial
range of our simulation. This quantity is the same as the Eulerian
2nd-order moment, giving the probability to have two particles within
a distance $r$.  We note that particle distribution recovers
uniformity at large scales very slowly, much slower than if particle
were distributed as Poisson point-like clusters for which $\langle
\rho^2\rangle\!-\!1 \!\propto\! r^{-3}\! \propto
\gamma_{r,\tau}^{9/5}$ (also shown in the inset for comparison).

We presented nu\-me\-ri\-cal evi\-den\-ce that at mo\-de\-ra\-te
Reynolds number the distribution of heavy particles at lengthscales
within the inertial range is fully described in terms of a
scale-dependent volume contraction rate $\gamma_{r,\tau}\propto\tau /
r^{5/3}$. However we expect that $\gamma_{r,\tau}\propto\tau /
r^{4/3}$ at sufficiently large Reynolds numbers ($Re_\lambda >
600$). State-of-the-art numerical simulations can not currently attain
these turbulent regimes, so that only experimental measurements of
particle distribution can help to confirm such a prediction. We have
seen that particle dynamics in the inertial range can be directly
related to the structure of the pressure field (and thus of
acceleration).  Characterizing the distribution of acceleration is
thus crucial to understand particle clusters, and conversely, particle
distribution could be used as an experimental tool to probe the
spatial structure of turbulent flows.

We acknowledge useful discussions with G.\ Boffetta, A.\ Celani and
A.\ Fouxon. This work has been partially supported by the EU under
contract HPRN-CT-2002-00300 and the Galileo programme on Trasport and
dispersion of impurities suspended in turbulent flows. Simulations
were performed at CINECA (Italy) and IDRIS (France) under the
HPC-Europa programme (RII3-CT-2003-506079).  Unprocessed data of this
study are freely available from http://cfd.cineca.it.


\begin{thebibliography}{18}
\expandafter\ifx\csname natexlab\endcsname\relax\def\natexlab#1{#1}\fi
\expandafter\ifx\csname bibnamefont\endcsname\relax
  \def\bibnamefont#1{#1}\fi
\expandafter\ifx\csname bibfnamefont\endcsname\relax
  \def\bibfnamefont#1{#1}\fi
\expandafter\ifx\csname citenamefont\endcsname\relax
  \def\citenamefont#1{#1}\fi
\expandafter\ifx\csname url\endcsname\relax
  \def\url#1{\texttt{#1}}\fi
\expandafter\ifx\csname urlprefix\endcsname\relax\def\urlprefix{URL }\fi
\providecommand{\bibinfo}[2]{#2}
\providecommand{\eprint}[2][]{\url{#2}}

\bibitem{cst98}
\bibinfo{author}{\bibfnamefont{C.}~\bibnamefont{Crowe}},
  \bibinfo{author}{\bibfnamefont{M.}~\bibnamefont{Sommerfeld}},
  \bibnamefont{and} \bibinfo{author}{\bibfnamefont{Y.}~\bibnamefont{Tsuji}},
  \emph{\bibinfo{title}{Multiphase Flows with Particles and Droplets}}
  (\bibinfo{publisher}{CRC Press}, \bibinfo{address}{New York},
  \bibinfo{year}{1998}).

\bibitem{pl01}
\bibinfo{author}{\bibfnamefont{I.}~\bibnamefont{de~Pater}} \bibnamefont{and}
  \bibinfo{author}{\bibfnamefont{J.}~\bibnamefont{Lissauer}},
  \emph{\bibinfo{title}{Planetary Science}} (\bibinfo{publisher}{Cambridge
  University Press}, \bibinfo{address}{Cambridge}, \bibinfo{year}{2001}).

\bibitem{ks01}
\bibinfo{author}{\bibfnamefont{A.}~\bibnamefont{Kostinski}}
\bibnamefont{and}
\bibinfo{author}{\bibfnamefont{R.}~\bibnamefont{Shaw}},
\bibinfo{journal}{J.\ Fluid Mech.} \textbf{\bibinfo{volume}{434}},
\bibinfo{pages}{389} (\bibinfo{year}{2001}).

\bibitem{ef94} \bibinfo{author}{\bibfnamefont{J.}~\bibnamefont{Eaton}}
\bibnamefont{and}
\bibinfo{author}{\bibfnamefont{J.}~\bibnamefont{Fessler}},
\bibinfo{journal}{Int.\ J.\ Multiphase Flow}
\textbf{\bibinfo{volume}{20}}, \bibinfo{pages}{169}
(\bibinfo{year}{1994}).

\bibitem{ck04}
\bibinfo{author}{\bibfnamefont{L.}~\bibnamefont{Collins}}
\bibnamefont{and}
\bibinfo{author}{\bibfnamefont{A.}~\bibnamefont{Keswani}},
\bibinfo{journal}{New J.\ Phys.} \textbf{\bibinfo{volume}{6}},
\bibinfo{pages}{119} (\bibinfo{year}{2004}).

\bibitem{ffs03}
\bibinfo{author}{\bibfnamefont{G.}~\bibnamefont{Falkovich}},
\bibinfo{author}{\bibfnamefont{A.}~\bibnamefont{Fouxon}},
\bibnamefont{and}
\bibinfo{author}{\bibfnamefont{M.}~\bibnamefont{Stepanov}}, in
\emph{\bibinfo{booktitle}{Sedimentation and Sediment Transport}},
edited by \bibinfo{editor}{\bibfnamefont{A.}~\bibnamefont{Gyr}}
\bibnamefont{and}
\bibinfo{editor}{\bibfnamefont{W.}~\bibnamefont{Kinzelbach}}
(\bibinfo{publisher}{Kluwer Academic Publishers},
\bibinfo{address}{Dordrecht}, \bibinfo{year}{2003}), pp.
\bibinfo{pages}{155--158}.

\bibitem{m87}
\bibinfo{author}{\bibfnamefont{M.}~\bibnamefont{Maxey}}, \bibinfo{journal}{J.\
  Fluid\ Mech.} \textbf{\bibinfo{volume}{174}}, \bibinfo{pages}{441}
  (\bibinfo{year}{1987}).

\bibitem{bff01}
  \bibinfo{author}{\bibfnamefont{E.}~\bibnamefont{Balkovsky}},
  \bibinfo{author}{\bibfnamefont{G.}~\bibnamefont{Falkovich}},
  \bibnamefont{and} \bibinfo{author}{\bibfnamefont{A.}~\bibnamefont{Fouxon}},
  \bibinfo{journal}{Phys.\ Rev.\ Lett.} \textbf{\bibinfo{volume}{86}},
  \bibinfo{pages}{2790} (\bibinfo{year}{2001}).

\bibitem{eklrs02}
  \bibinfo{author}{\bibfnamefont{T.}~\bibnamefont{Elperin}},
  \bibinfo{author}{\bibfnamefont{N.}~\bibnamefont{Kleeorin}},
  \bibinfo{author}{\bibfnamefont{V.}~\bibnamefont{L'vov}},
  \bibinfo{author}{\bibfnamefont{I.}~\bibnamefont{Rogachevskii}},
  \bibnamefont{and}
  \bibinfo{author}{\bibfnamefont{D.}~\bibnamefont{Sokoloff}},
  \bibinfo{journal}{Phys.\ Rev.} E \textbf{\bibinfo{volume}{66}},
  \bibinfo{pages}{036302} (\bibinfo{year}{2002}).

\bibitem{noi1} \bibinfo{author}{\bibfnamefont{J.}~\bibnamefont{Bec}},
\bibinfo{author}{\bibfnamefont{L.}~\bibnamefont{Biferale}},
\bibinfo{author}{\bibfnamefont{G.}~\bibnamefont{Boffetta}},
\bibinfo{author}{\bibfnamefont{A.}~\bibnamefont{Celani}},
\bibinfo{author}{\bibfnamefont{M.}~\bibnamefont{Cencini}},
\bibinfo{author}{\bibfnamefont{A.}~\bibnamefont{Lanotte}},
\bibinfo{author}{\bibfnamefont{S.}~\bibnamefont{Musacchio}},
\bibnamefont{and}
\bibinfo{author}{\bibfnamefont{F.}~\bibnamefont{Toschi}},
\bibinfo{journal}{J.\ Fluid\ Mech.} \textbf{\bibinfo{volume}{550}},
\bibinfo{pages}{349} (\bibinfo{year}{2006});
\bibinfo{author}{\bibfnamefont{M.}~\bibnamefont{Cencini}},
\bibinfo{author}{\bibfnamefont{J.}~\bibnamefont{Bec}},
\bibinfo{author}{\bibfnamefont{L.}~\bibnamefont{Biferale}},
\bibinfo{author}{\bibfnamefont{G.}~\bibnamefont{Boffetta}},
\bibinfo{author}{\bibfnamefont{A.}~\bibnamefont{Celani}},
\bibinfo{author}{\bibfnamefont{A.}~\bibnamefont{Lanotte}},
\bibinfo{author}{\bibfnamefont{S.}~\bibnamefont{Musacchio}},
\bibnamefont{and}
\bibinfo{author}{\bibfnamefont{F.}~\bibnamefont{Toschi}},
\bibinfo{journal}{Journ.\ Turb.} \textbf{\bibinfo{volume}{7}},
\bibinfo{pages}{1} (\bibinfo{year}{2006}).



\bibitem{ekr96}
  \bibinfo{author}{\bibfnamefont{T.}~\bibnamefont{Elperin}},
  \bibinfo{author}{\bibfnamefont{N.}~\bibnamefont{Kleeorin}},
  \bibnamefont{and}
  \bibinfo{author}{\bibfnamefont{I.}~\bibnamefont{Rogachevskii}}, ,
  \bibinfo{journal}{Phys.\ Rev.\ Lett.} \textbf{\bibinfo{volume}{77}},
  \bibinfo{pages}{5373} (\bibinfo{year}{1996}).

\bibitem{bch06} \bibinfo{author}{\bibfnamefont{J.}~\bibnamefont{Bec}},
\bibinfo{author}{\bibfnamefont{M.}~\bibnamefont{Cencini}},
\bibnamefont{and}
\bibinfo{author}{\bibfnamefont{R.}~\bibnamefont{Hillerbrand}}, Physica
D, in press (2006).

\bibitem{cgv06}
\bibinfo{author}{\bibfnamefont{L.}~\bibnamefont{Chen}},
  \bibinfo{author}{\bibfnamefont{S.}~\bibnamefont{Goto}}, \bibnamefont{and}
  \bibinfo{author}{\bibfnamefont{J.}~\bibnamefont{Vassilicos}},
  \bibinfo{journal}{J.\ Fluid\ Mech.} \textbf{\bibinfo{volume}{553}},
  \bibinfo{pages}{143} (\bibinfo{year}{2006}).

\bibitem{achl02}
\bibinfo{author}{\bibfnamefont{A.}~\bibfnamefont{Aliseda}},
\bibinfo{author}{\bibfnamefont{A.}~\bibfnamefont{Cartellier}},
\bibinfo{author}{\bibfnamefont{F.}~\bibfnamefont{Hainaux}},
\bibnamefont{and}
\bibinfo{author}{\bibfnamefont{J.C.}~\bibnamefont{Lasheras}},
\bibinfo{journal}{J.\ Fluid\ Mech.} \textbf{\bibinfo{volume}{468}},
\bibinfo{pages}{77} (\bibinfo{year}{2002}).

\bibitem{bdg04}
\bibinfo{author}{\bibfnamefont{G.}~\bibnamefont{Boffetta}},
  \bibinfo{author}{\bibfnamefont{F.}~\bibnamefont{De~Lillo}}, \bibnamefont{and}
  \bibinfo{author}{\bibfnamefont{A.}~\bibnamefont{Gamba}},
  \bibinfo{journal}{Phys.\ Fluids} \textbf{\bibinfo{volume}{16}},
  \bibinfo{pages}{L20} (\bibinfo{year}{2004}).

\bibitem{bgh04}
  \bibinfo{author}{\bibfnamefont{J.}~\bibnamefont{Bec}},
  \bibinfo{author}{\bibfnamefont{K.}~\bibnamefont{Gaw\c{e}dzki}},
  \bibnamefont{and}
  \bibinfo{author}{\bibfnamefont{P.}~\bibnamefont{Horvai}},
  \bibinfo{journal}{Phys.\ Rev.\ Lett.} \textbf{\bibinfo{volume}{92}},
  \bibinfo{pages}{224501} (\bibinfo{year}{2004}).

\bibitem{bch06b}
\bibinfo{author}{\bibfnamefont{J.}~\bibnamefont{Bec}},
  \bibinfo{author}{\bibfnamefont{M.}~\bibnamefont{Cencini}}, \bibnamefont{and}
  \bibinfo{author}{\bibfnamefont{R.}~\bibnamefont{Hillerbrand}}
  (\bibinfo{year}{2006}{\natexlab{c}}), \bibinfo{note}{preprint
  nlin.CD/0606038}.

\bibitem{wm05}
\bibinfo{author}{\bibfnamefont{M.}~\bibnamefont{Wilkinson}}
\bibnamefont{and}
\bibinfo{author}{\bibfnamefont{B.}~\bibnamefont{Mehlig}},
\bibinfo{journal}{Europhys.\ Lett.} \textbf{\bibinfo{volume}{71}},
\bibinfo{pages}{186} (\bibinfo{year}{2005}).

\bibitem{gf01-ti03} T.~Gotoh and D.~Fukayama, Phys.\ Rev.\
Lett.~\textbf{86}, 3775 (2001); Y.~Tsuji and T.~Ishihara, Phys.\
Rev.~E~\textbf{68}, 026309 (2003).

\bibitem{t75} H.\ Tennekes, J.\ Fluid Mech.~\textbf{67}, 561 (1975).
\end{thebibliography}
\end{document}